\input{aipcheck}

\documentclass[
    final            
  ]{aipproc}

\layoutstyle{8x11double}

\usepackage{amssymb,amsmath}
\usepackage{cancel}
\usepackage{bbm} 

\def\MPl{\ensuremath{M_\mathrm{Pl}}}    
\def\MSt{\ensuremath{M_\mathrm{St}}}    
\def\D{\ensuremath{\mathrm{d}}}                 
\newcommand{\antiD}[1]{\ensuremath{\overline{\mathrm{D}#1}}} 
\DeclareMathOperator{\re}{Re}

\DeclareMathOperator{\Tr}{Tr}

\begin{document}

\title{$\boldsymbol{\log(M_\mathrm{Pl}/m_{3/2})}$}


\author{Oscar Loaiza-Brito}{%
  address={Physikalisches Institut, Universit\"at Bonn\\
  Nussallee 12, D-53115 Bonn, Germany}
}

\author{Johannes Martin}{%
  address={Physikalisches Institut, Universit\"at Bonn\\
  Nussallee 12, D-53115 Bonn, Germany}
}

\author{Hans Peter Nilles}{%
  address={Physikalisches Institut, Universit\"at Bonn\\
  Nussallee 12, D-53115 Bonn, Germany}
}

\author{Michael Ratz}{
  address={Physikalisches Institut, Universit\"at Bonn\\
  Nussallee 12, D-53115 Bonn, Germany}
}

\begin{abstract}
  Flux compactifications of string theory seem to require the presence
  of a fine-tuned constant in the superpotential. We discuss a scheme
  where this constant is replaced by a dynamical quantity which we
  argue to be a `continuous Chern--Simons term'. In such a scheme, the
  gaugino condensate generates the hierarchically small scale of
  supersymmetry breakdown rather than adjusting its size to a
  constant. A crucial ingredient is the appearance of the
  hierarchically small quantity $\exp(-\langle X\rangle)$ which
  corresponds to the scale of gaugino condensation. Under rather
  general circumstances, this leads to a scenario of moduli
  stabilization, which is endowed with a hierarchy between the mass of
  the lightest modulus, the gravitino mass and the scale of the soft
  terms, $m_\mathrm{modulus}\sim \langle X\rangle\,m_{3/2}\sim \langle
  X\rangle^2\,m_\mathrm{soft}$. The `little hierarchy' $\langle
  X\rangle$ is given by the logarithm of the ratio of the Planck scale
  and the gravitino mass, $\langle
  X\rangle\sim\log(\MPl/m_{3/2})\sim4\pi^2$. This exhibits a new
  mediation scheme of supersymmetry breakdown, called mirage mediation. We highlight the special
  properties of the scheme, and their consequences for phenomenology
  and cosmology.
\end{abstract}

\maketitle


\section{Introduction}

Superstring theories are the most attractive candidates for a unified
description of all observed phenomena. They provide all structures necessary to
accommodate the matter content of the standard model as well as all known
interactions. However, a commonly accepted stringy extension of the standard
model has not yet emerged. Apart from the obvious problem to obtain the correct
spectrum there are further, more fundamental questions, which have to be answered
if one wants to relate superstring theory to observation. These questions
include:
\renewcommand\labelenumi{(\roman{enumi})}
\begin{enumerate}
 \item Why is the scale of weak interactions so much lower than the scale of
  gravity?\label{q1}
 \item Why do we observe four space-time dimensions?\label{q2}
 \item Why do we live in de Sitter (or Minkowski) space?\label{q3}
\end{enumerate}
The first question concerns the appearance of the weak scale $m_\mathrm{weak}$
while string and Planck scale, $\MSt$ and $\MPl$, are of similar size, and
$m_\mathrm{weak}\ll \MPl$. To address the second question, one usually confines
oneself to the problem of finding a self-consistent compactification from ten to
four dimensions. This includes, in particular, the stabilization of the moduli,
which parametrize the size and shape of the internal space. The last question is
highly non-trivial since string compactifications admit anti-de Sitter (adS)
minima, i.e.\ vacua with negative vacuum energy. It is challenging to understand
in such a framework why the vacuum chosen by nature has positive (or zero)
energy.

These questions are not unrelated. In four dimensions, hierarchically small
scales can be obtained by dimensional transmutation. The conventional approach
to address the hierarchy problem consists in generating a hierarchically small
scale of supersymmetry (SUSY) breakdown by a non-perturbative effect, such as a
gaugino condensate \cite{Nilles:1982ik}. This leads to the appearance of the
scale $M_{\cancel{\mathrm{SUSY}}}\sim\MSt\exp(-X)$ with $X$ being a moderately
large field-dependent quantity. Once SUSY is broken, the moduli get a
non-trivial potential, which might result in their stabilization. However, it is
rather difficult to obtain consistent scenarios where a stabilization of all
moduli occurs at realistic values. Furthermore, in this picture, one would
encounter a situation where the mass of (most of) the moduli is of the order of
the weak scale. It is, however, known that such moduli masses lead to severe
problems for cosmology.

More recently, the picture has changed due to significant progress in
understanding the role of fluxes for moduli stabilization
\cite{Giddings:2001yu}. The main new feature is that some of the moduli can be
fixed at realistic values while attaining masses of the order $\MSt$. 

Some important aspects of the `flux compactification' scheme are nicely
illustrated by the toy example of KKLT \cite{Kachru:2003aw} in type IIB string
theory. Here, in a first step the complex structure moduli ($Z^a$) and the
dilaton ($S$) get stabilized by fluxes. This results in the appearance of a
(fine-tuned) constant in the superpotential which, at this stage, breaks SUSY
with the scale of SUSY breaking being set by the size of the constant. In the
second step, a gaugino condensate is included, which adjusts its size to this
constant, thereby fixing the K\"ahler modulus ($T$) and restoring SUSY. At this
stage, the vacuum energy is negative. This has to be rectified in the third step
where an (ad-hoc) `uplifting' is introduced, which renders the vacuum energy
positive, and then (again) breaks SUSY.  Clearly, in the context of flux
compactifications one usually loses the explanation of the hierarchy
$m_\mathrm{weak}\sim m_\mathrm{soft}\ll \MPl$.  This hierarchy is now related to
the appearance of a small (quantized) constant in the superpotential, which
requires a severe fine-tuning.

In Sec.~2, we propose a modification of the `flux compactification'
scheme where the positive features of the latter are retained while
(some of) the problematic aspects are avoided. The main novelty is
that the constant in the superpotential gets replaced by a dynamical
quantity. This means that after the first step of moduli stabilization
SUSY is unbroken and the superpotential vanishes, leading to zero
vacuum energy at this stage. As a consequence, the non-perturbative
effect (gaugino condensate) sets the scale of SUSY breaking rather
than adjusting to a constant. We argue that the above-mentioned
dynamical quantity should be given by the continuous, i.e.\ 
non-quantized, part of the Chern--Simons term appearing in dimensional
reduction from ten to four dimensions.

In Sec.~3, we discuss the phenomenological consequences of the
appearance of the small quantity $\exp(-X)$, which leads under rather
general circumstances to the `little mass hierarchy'
\begin{equation}
 m_\mathrm{moduli}
 ~\sim~
 \langle X\rangle\,m_{3/2}
 ~\sim~
 \langle X\rangle^2\, m_\mathrm{soft}\;,
\end{equation}
where $\langle X\rangle\sim\log(\MPl/m_{3/2})\sim4\pi^2$. As we shall
see, this results in a scheme with distinct properties. These
properties solve (or, at least, help to solve) several problems of
supersymmetric extensions of the standard model.

\section{Avoiding quantized constants}
\label{sec:ChernSimons}

The scenario of hidden sector gaugino condensation yields a very plausible
explanation of the hierarchy $m_\mathrm{weak}\ll \MPl$. Here, strong dynamics
leads to the non-trivial expectation value of the gaugino bilinear
\cite{Nilles:1982ik}, $\langle\lambda\lambda\rangle= \Lambda^3$, where $\Lambda$
is of the order of the renormalization group (RG) invariant scale,
\begin{equation}\label{eq:RGinvariant}
 \Lambda ~\sim~ \mu\exp\{-1/[b_0\,g^2(\mu)]\}~\ll~\MPl\;.
\end{equation}
$b_0$ is the coefficient of the $\beta$-function. This strong dynamics triggers
a breakdown of SUSY that is parametrized by the gravitino mass
\begin{equation}\label{eq:MSUSY}
 m_{3/2}~\sim~\frac{\Lambda^3}{\MPl^2}
 \quad\text{and}\quad
 M_{\cancel{\mathrm{SUSY}}}~\simeq~\sqrt{m_{3/2}\MPl}\;.
\end{equation}
Notice that SUSY breakdown requires non-trivial gauge-kinetic function
\cite{Ferrara:1982qs},
\begin{equation}
 F_i~=~\exp(-K)\,D_iW + f_i\, \langle\lambda\lambda\rangle + \dots\;.
\end{equation}
In other words, the gauge coupling has to be field-dependent, $g^{-2}=f$.
It is further possible to include the gaugino condensate in the superpotential
\cite{Dine:1985rz},
\begin{equation}
 W~=~W_\mathrm{perturbative}+C\,\exp(-a\,f)\;.
\end{equation}

Early attempts to incorporate gaugino condensation in (heterotic) string theory
\cite{Derendinger:1985kk,Dine:1985rz} revealed the importance of a background
flux of the field strength $H$ of the 2-index antisymmetric tensor field $B$ to
avoid a run-away behavior of the dilaton field. This is obvious  from the
`perfect square' structure of supergravity \cite{Chamseddine:1980cp},
\begin{equation}\label{eq:perfectsquare}
 S_\mathrm{SUGRA}~\supset~
 \left(H - \alpha^\prime \,\langle\lambda\lambda\rangle\right)^2
 \;
\end{equation}
Here, $H$ is the 3-form field strength of the two-index antisymmetric tensor
field $B$ that appears in the 10d supergravity multiplet. It is important to
note that the naive field strength $H_\mathrm{naive}=\D B$ has to be amended
by Chern--Simons terms \cite{Chamseddine:1980cp,Green:1984sg},
\begin{equation}
 H~=~
 \D B
 -\frac{1}{\sqrt{2}}\left(\omega^{(\mathrm{YM})}-\omega^{(\mathrm{L})}\right)
 \;,
\end{equation}
where the Yang--Mills Chern--Simons term is given by
\begin{eqnarray}\label{eq:ChernSimons}
 \omega^{(\mathrm{YM})}_{MNP}
 & = &
 \Tr\left(A_{[M}\,F_{NP]}-\frac{2}{3}A_{[M}\,A_N\,A_{P]}\right)\;,
\end{eqnarray}
and an analogous expression exists for the Lorentz Chern--Simons term.

The `perfect square structure' \eqref{eq:perfectsquare} leads to the
possibility that the flux $H$ stabilizes the gaugino condensate (or
vice versa) \cite{Derendinger:1985kk,Dine:1985rz}.  However,
generically not all moduli are stabilized. Moreover, the gaugino
condensate can no longer account for the hierarchy, since it balances
the value of the quantized $H$~\cite{Rohm:1985jv}.  If one were to set
$H=0$, the latter problem would not arise. However, in such a scenario
a non-trivial value of the gaugino condensate would now lead to a
vacuum energy of order $\Lambda^6/\MPl^2$, which is inconsistent with
observation. One would therefore need a small, non-quantized piece of
$H$ to conspire with the gaugino condensate such that the vacuum
energy (almost) vanishes.

Our statements rely on the non-quantized nature of the Chern--Simons
terms. The quantization of $H$ was shown in~\cite{Rohm:1985jv} for the
case $H=\D B$ and did not take into account the appearance of the
Chern--Simons terms.
\footnote{When compactifying on a compact space $K$ with
  $\pi_i(K)=\mathbbm{Z}_n$ there are fractional contributions to $H$,
  $\delta H =1/n$ \cite{Rohm:1985jv}. It has been argued that this
  might be used stabilize the dilaton \cite{Gukov:2003cy}.  However,
  we find it difficult to imagine that $1/n$ could explain the
  hierarchy between string and weak scale.}
If $\D B=0\,$ to leading order, it was argued in
  \cite{Derendinger:1985cv} that the cancellation should take place
  between the gaugino condensate and the Chern--Simons terms, which
  avoid the quantization constraint.
  
  Let us spell out these arguments in more detail.  We are interested
  in the $A_mA_nA_p$ part of the Chern--Simons term
  \eqref{eq:ChernSimons} where $m$, $n$, $p$ are indices w.r.t.\ the
  internal dimensions. As is well known, those internal components of
  the gauge fields can come in two different types:
\begin{enumerate}
\item On the one hand, the `discrete Wilson lines'
  \cite{Ibanez:1986tp} correspond to quantized background values of
  $A_n^a\mathsf{t}_a$ ($\mathsf{t}_a$ denotes the generator) with
  support on non-contractible loops in the internal space.  They take
  values in the adjoint representation so that switching them on does
  not reduce the rank of the gauge group. To understand the
  quantization of the `discrete Wilson lines' $A_n^a\mathsf{t}_a$,
  observe that an adjoint expectation value does not break the U(1)
  generated by $\mathsf{t}_a$.  Consequently, the expectation values
  of `discrete Wilson lines' are quantized to ensure that the
  zero-modes living on the above-mentioned non-contractible loops are
  single-valued.
\item `Continuous Wilson lines' \cite{Ibanez:1987xa}, on the other
  hand, transform in the coset of the gauge group (which is present
  before they are switched on). A generic expectation value of a
  `continuous Wilson line' does reduce the rank. Since the U(1)
  generated by $\mathsf{t}_a$ is (generically) broken, there is no
  quantization constraint for `continuous Wilson lines'. Let us
  finally mention that in orbifold compactifications
  \cite{Dixon:1985jw,Dixon:1986jc} `continuous Wilson lines' emerge
  from the untwisted sector, and can be interpreted as `matter fields'
  in the massless spectrum \cite{Font:1988tp}.
\end{enumerate}

It is now clear that the trilinear term of three continuous Wilson
lines can attain arbitrary values, and does in particular not suffer
from quantization. It is precisely this term, which can adjust to a
gaugino condensate, thus cancelling the corresponding potential
energy. In the following, we will refer to such terms as `continuous
Chern--Simons terms'. We observe that $\omega^{(\mathrm{YM})}$ and
$\langle\lambda\lambda\rangle$ are both `$\alpha^\prime$ corrections',
thus suggesting their alignment without involving the quantized
flux.

\begin{figure}[h]
  \includegraphics[height=.18\textheight]{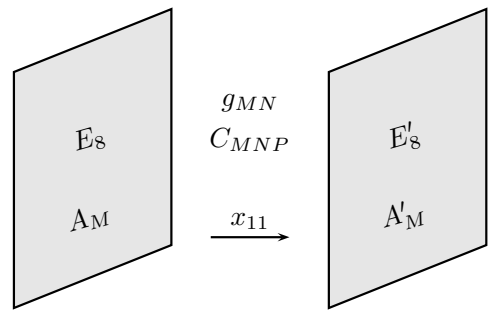}
  \caption{M-theory set-up. $C$ lives in the bulk whereas both the gaugino
  condensate and the (continuous) Chern--Simons terms live on the branes.}
  \label{fig:Mtheory}
\end{figure}

Further support for the cancellation between the gaugino condensate
and the continuous Chern--Simons term is provided within the framework
of heterotic M-theory of Ho{\u{r}}ava and Witten
\cite{Horava:1995qa,Horava:1996ma}. In M-theory, gravity lives in the
11d bulk whereas the gauge fields reside on the two 10d boundaries.
In the 11d bulk supergravity multiplet we find a 3-index tensor field
$C_{MNP}$ with the four index field strength $G=\D
C+\text{Chern--Simons terms}$.  Dimensionally reducing to 10
dimensions one finds that $B_{MN}$ descends form $C_{MN,11}$ with the
corresponding relation between $H$ and $G$.
It is now clear that the Chern--Simons terms are located on the boundaries,
\begin{eqnarray}
 G~=~\D C+\alpha^\prime \sum_i\delta(x_{11}-x_{11}^i)\,
\left(\omega_i^{(\mathrm{YM})}-\frac{1}{2}\omega_i^{(\mathrm{L})}\right)
\end{eqnarray}
with $\D G=\Tr F_1^2 +\Tr F_2^2 - \Tr R^2$ where $F_1$ and $F_2$ represent the
field strengths of the two $\mathrm{E}_8$ factors.
It has further been shown in Ref.~\cite{Horava:1996vs} that the
perfect square structure between the flux $G$ and the gaugino
condensate generalizes to this case. Since the gauginos are also
fields confined to the boundaries, we consider this as a further
argument for a cancellation between the gaugino condensate and the
Chern--Simons terms \cite{Nilles:1997cm,Nilles:1998sx}, while the
quantized bulk contribution $\D C$ should not contribute to this
cancellation, thus avoiding any known quantization constraint. We
sketch this local cancellation in Fig.~\ref{fig:Mtheory}.

Let us now summarize the outcome of the discussion so far. We have
argued that the appearance of quantized constants in the
superpotential should be avoided in order to explain the hierarchy
between the Planck and the weak scale by a natural mechanism such as
gaugino condensation. We have also discussed that the field strength
$H$ and an expectation value of the gaugino bilinear combine into a
perfect square \eqref{eq:perfectsquare}, and that $H$ contains both
quantized and continuous parts. This leads to the possibility that the
continuous Chern--Simons term cancels the contribution of the gaugino
condensate to the vacuum energy.

We have now provided all the ingredients of a modification of the
`flux compactification' scheme, which can be outlined as follows: In a
first step, before the inclusion of gaugino condensate and continuous
Chern--Simons terms, the moduli are fixed by quantized 3-form fluxes
without breaking SUSY---i.e., $D_iW=0$ for all fields and $W=0$ in the
vacuum. The inclusion of the gaugino condensate, in the second step,
leads to SUSY breakdown. As a consequence,
$M_{\cancel{\mathrm{SUSY}}}$ is now explained by dimensional
transmutation (cf.\ Eq.~\eqref{eq:MSUSY}). At this stage, the vacuum
energy is of order $M_{\cancel{\mathrm{SUSY}}}^2$, i.e.\ unacceptably
large. This vacuum energy can be compensated for by the continuous
Chern--Simons term. A more detailed discussion and examples will be
presented elsewhere \cite{Loaiza2005:pr}.

\section{Mirage Mediation}
\label{sec:MirageMediation}

Let us now investigate the phenomenological properties of the scheme where all
but one modulus are fixed by fluxes and the last one gets stabilized through
non-perturbative effects such as a gaugino condensate. 

\subsection{General structure of the scheme}

The stabilization of the last modulus is described by the following
model-independent structure of the (effective) superpotential:
\begin{eqnarray}
 W~=~A + C\,\exp (-X)\;.
\end{eqnarray}
Here, $A$ and $X$ represent vacuum expectation values of
field-dependent quantities, and $C\sim\MPl^3$. The value of $A$ has to
be small compared to the string/Planck scale, which can be achieved
either through a natural mechanism, or through an explicit
fine-tuning. The gravitino mass $m_{3/2}$ will appear as
\begin{eqnarray}\label{eq:W}
 W~\simeq~\frac{m_{3/2}}{\MPl}\times\MPl^3\;,
\end{eqnarray}
where the Planck scale is assumed to be of order of the string scale. The
K\"ahler potential is (up to a constant)
\begin{equation}
 K~=~-n\,\log\left(X+\overline{X}\right)+\dots
\end{equation}
where the omission denotes the K\"ahler potential for matter fields, and $n$ is
an order one constant. The scalar potential is given by
\begin{equation}\label{eq:V}
 V~=~e^K\,\left[K^{\alpha\bar \beta}(D_\alpha W)(D_{\bar \beta}\overline{W})
 -3|W|^2\right]\;,
\end{equation}
with the K\"ahler derivative $D_\alpha W=\partial_\alpha W+K_\alpha W$, and
where we set $\MPl=1$.
The minimum occurs for $D_XW\sim0$, i.e.\ by Eq.~\eqref{eq:W} for
\begin{eqnarray}\label{eq:X4pi}
 X~\sim~\log(\MPl/m_{3/2})~\sim~4\pi^2\;,
\end{eqnarray}
where we indicate the approximate numerical value of the logarithm of the
hierarchy between $m_{3/2}$ and $\MPl$.

To arrive at zero vacuum energy, we have to arrange a cancellation between the
terms in the brackets of \eqref{eq:V} which are, when multiplied by $e^K$, both
of the order of the square of the gravitino mass. We  therefore have
\begin{equation}
 V''|_{X=\langle X\rangle}~ \sim ~ m_{3/2}^2\;,
\end{equation}
where the prime indicates the derivative w.r.t.\ $\re X$, and $\langle X\rangle$
denotes the position of the minimum. To evaluate the physical mass, one has to
make sure that the kinetic term of the fluctuations $\delta X$ around the
minimum is canonical,
\begin{eqnarray}
 K & = &
 -n\,\log\left(\langle X\rangle+\langle\overline{X}\rangle\right)
 -\frac{n}{\langle X\rangle+\langle\overline{X}\rangle}\,\left(\delta X+\delta\overline{X}\right)
 \nonumber\\
 & &{}
 +\frac{n}{\left(\langle X\rangle+\langle\overline{X}\rangle\right)^2}\,
        \left(\delta X\,\delta\overline{X}\right)+\dots\;.
 \label{eq:canonical}
\end{eqnarray}
This amounts to a rescaling $\delta X\to\delta X_\mathrm{can}=\delta
X\times(\sqrt{n}/\re \langle X\rangle)$. In particular, one finds for
the physical mass of the modulus (taking $\langle X\rangle$ to be real
and positive)
\begin{equation}\label{eq:enhancedMass}
 m_X~\sim~m_{3/2}\times\langle X\rangle\;.
\end{equation}
This enhancement of moduli masses is known to be a rather generic feature of the
non-perturbative moduli stabilization mechanisms
\cite{Banks:1994sg,Banks:1996ea}. We have sharpened the statement, and in
particular shown that this enhancement occurs when (i) the K\"ahler potential
for $X$ is logarithmic, and (ii) the dependence of the superpotential contains
the exponential term such that $\exp(-\langle X\rangle)\sim \Lambda^3$ (with
$\Lambda$ as in Eqs.~\eqref{eq:RGinvariant} and \eqref{eq:MSUSY}). Using
\eqref{eq:X4pi} we can recast \eqref{eq:enhancedMass} as
\begin{equation}\label{eq:enhancedMass2}
 m_X~\sim~m_{3/2}\times\log(\MPl/m_{3/2})~\sim~m_{3/2}\times 4\pi^2\;.
\end{equation}

\subsection{An example}

Let us now discuss specifically the outcome in the simple model of KKLT
\cite{Kachru:2003aw} with matter fields on D7-branes as analyzed in
\cite{Choi:2004sx,Choi:2005ge}. We concentrate on the case with the dilaton $S$,
a  K\"{a}hler modulus $T$ and complex structure moduli $Z_\alpha$. Matter
superfields are denoted by $Q_7$. We assume to be in a region of large $S$ and
$T$. Let us start with the D7-system
\cite{Camara:2003ku,Grana:2003ek,Grimm:2004uq,Lust:2004cx,Lust:2004fi,Camara:2004jj,Ibanez:2004iv,Jockers:2004yj,Lust:2004dn}.  
The K\"{a}hler potential is assumed to be
\begin{eqnarray}
 K & = & - \log (S + \overline{S} - |Q_7|^2   ) - 3 \log (T + \overline{T} )
 \nonumber\\
 & & {}
 + \tilde{K}(Z_\alpha, \overline{Z}_\alpha)\;,\label{kaehler-IIA}
\end{eqnarray}
where  $Q_7$ denote matter multiplets on the  D7 branes. The gauge kinetic
function is
\begin{equation}\label{gauge-IIA}
 f_7~ =~ T
\end{equation}
for gauge bosons on the D7 branes. The inclusion of fluxes leads to a
superpotential for the moduli $S$ and $Z_\alpha$ \cite{Kachru:2003aw}. As a
consequence,  one can eliminate (`integrate out') these fields
\cite{Choi:2004sx,deAlwis:2005tg,deAlwis:2005tf}. This leads to an effective
superpotential which is given by
\begin{equation}
 W~=~W(S, Z_\alpha) + C\, \exp (-aT) + W(Q_7)\;,
\end{equation}
where $C\sim\MPl^3$ and $a$ are constants. The term
$C\exp(-aT)$ represents gaugino condensation on the D7-branes. When
analyzing the potential, we look for minima where the $Q_7$ scalars
(and therefore $W(Q_7)$ as well) do not receive non-trivial vacuum
expectation values. Extremizing the scalar potential w.r.t.\ $T$ leads
to an anti-de Sitter vacuum with energy $\sim |W(S,
Z_\alpha)/\MPl|^2$. To render this vacuum realistic, one introduces an
(ad hoc) uplifting, which may be parametrized as
\begin{equation}\label{eq:Vlift}
 V_\mathrm{lift}~=~\frac{D}{\left(T+\overline{T}\right)^{n_T}}\;.
\end{equation}
By tuning $D$ it is possible to obtain local de Sitter vacua with energy consistent
with observation. The relevant scales appearing in such a vacuum have been
calculated in \cite{Choi:2005ge}, and they are given by:
\begin{eqnarray}
 \MSt &\sim& 5\times 10^{17} \,\mathrm{GeV}\;,\nonumber \\
 1/R &\sim& 10^{17} \,\mathrm{GeV}\;, \nonumber \\
 m_{Z,S} &\sim& \frac{1}{\MSt^2\,R^3} ~\sim~ 10^{16} \,\mathrm{GeV}\;,
 \nonumber \\
 \Lambda_{GC}&=& \MSt\,e^{-\langle aT\rangle/3} ~\sim~ 10^{13}\,\mathrm{GeV}\;,
 \nonumber \\
 M_{\antiD3} &\sim& e^{A_\mathrm{min}}\,\MSt ~\sim~
 10^{11} \,\mathrm{GeV}\;,
 \nonumber \\
 m_T &\sim& \langle aT\rangle m_{3/2} ~\sim~ 10^6\,\mathrm{GeV}\;,
 \nonumber \\
 m_{3/2} &\sim& \frac{1}{\MSt^2\,R^3}\left(\frac{G_{(0,3)}}{G_{(2,1)}}\right)
 ~\sim~ 10^4 \, \mathrm{GeV}\;,\nonumber \\
 m_\mathrm{soft} &\sim& m_\mathrm{weak}~\sim~
 \frac{m_{3/2}}{\langle aT\rangle}~\sim~ 10^2 \,\mathrm{GeV}\;,
 \label{eq:massscales}
\end{eqnarray}
where $\Lambda_{GC}$ is the dynamical scale of D7 gaugino condensation,
$M_{\antiD3}$ is the red-shifted cutoff scale on $\antiD3$, $e^{-\langle
aT\rangle}\sim m_{3/2}/\MSt$ and $e^{A_{min}}\sim \sqrt{m_{3/2}/\MSt}$.
$G_{(0,3)}$ and $G_{(2,1)}$ denote the (0,3) and (2,1) components of the flux
$G$. As is obvious from the above expressions, the SUSY breaking component
$G_{(0,3)}$ is substantially suppressed against $G_{(2,1)}$, which preserves
SUSY. In this case we have $X=aT$ and it will have a vacuum expectation value of
order $\langle X\rangle\sim \log(\MPl/m_{3/2})\sim 4\pi^2$ as we have discussed 
earlier.

\subsection{\cancel{SUSY} mediation}

SUSY is broken by the uplifting (cf.\ Eq.~\eqref{eq:Vlift}). To describe the
SUSY breakdown in the usual language, one attributes the associated $F$-term
expectation value to the so-called chiral compensator field $\widetilde{C}$
\cite{Choi:2005ge}. To see what this means, recall the usual supergravity
relation (in the absence of $D$-term expectation values)
\begin{equation}\label{eq:m32F}
 m_{3/2}^2~\sim~\sum_i\frac{F_i^2}{M_\mathrm{P}^2}\;,
\end{equation}
where the sum extends over the $F$-term expectation values of all chiral fields.
Here, the dominant $F$-term is the one of the chiral compensator which is
adjusted such that \eqref{eq:m32F} holds. On the other hand, the $F$-term of the
$T$-modulus is suppressed (cf.\ Eq.~\eqref{eq:massscales}).

Let us now explain how the suppressed $F_T$ term emerges. Before uplifting,
$F_T$ vanishes, and $T$ is stabilized with a mass $\sim\langle
aT\rangle\,m_{3/2}$ where $m_{3/2}=e^{K/2}|W|$ is the (adS) gravitino mass.
Uplifting does (practically) not change $m_{3/2}$ but depends on $T$ (cf.\
Eq.~\eqref{eq:Vlift}). As a consequence $T$ is slightly moved against its
original minimum after uplifting. The shift in $T$ is easily calculated in terms
of the canonically normalized fluctuations around the minimum (cf.\
Eq.~\eqref{eq:canonical})
\[
 \frac{\D}{\D \delta \overline{X}_\mathrm{can}}m_T^2 |\delta X_\mathrm{can}|^2 
 ~\stackrel{!}{=}~ -\frac{\D}{\D \delta \overline{X}_\mathrm{can}}V_\mathrm{lift}
 ~\sim~m_{3/2}^2\,\MPl^2\;,
\]
where we used in the last relation that $V_\mathrm{lift}$ is tuned such as to
cancel the negative energy of the adS minimum,
$V_\mathrm{adS}=-3|W|^2 e^K=-3\,m_{3/2}^2\MPl^2$. This leads to
$\delta X_\mathrm{can}\sim\MPl/\langle X\rangle^2$ so that in the shifted de
Sitter minimum  
\begin{equation}
 F_T^2~\sim~\frac{m_{3/2}^2\MPl^2}{\langle X\rangle^2}\;.
\end{equation}
This implies that soft terms induced by $F_T$ are suppressed against $m_{3/2}$
by a factor $\sim\langle X\rangle$. In particular, it is the same factor
$\langle X\rangle$, which both enhances the modulus mass and suppresses the
modulus $F$-term.

Hence, the `gravity mediated' (or `modulus mediated') soft terms,
being controlled by $F_T/T$, are suppressed against the gravitino mass, with the
suppression factor $(F_T/T)/m_{3/2}\sim F_T/F_{\widetilde{C}}\sim1/\langle
X\rangle$.  This suppression is comparable to a loop-factor, and therefore
anomaly mediation \cite{Randall:1998uk,Giudice:1998xp} becomes competitive. As a
consequence, the soft mass terms receive comparable contributions both from the
$F$-term of the $T$-modulus (`gravity mediation') and from the super-conformal
anomaly (`anomaly mediation'). In general, one might hence expect that such a
mix is a generic property of `sequestered' models where the communication of
SUSY breakdown can be more suppressed than by the Planck scale. We will call
this scheme `\textbf{mirage mediation}' in the following.

Let us emphasize the two features of mirage mediation that are most important
for cosmology and phenomenology:
\begin{itemize}
 \item The mass of $T$ is governed by SUSY breakdown. Yet this mass is enhanced
  with respect to the value of the gravitino mass (cf.\
  Eq.~\eqref{eq:enhancedMass}), $m_T = \langle X\rangle  m_{3/2}\sim 4\pi^2
  m_{3/2}$, and thus becomes quite heavy. 
 \item The soft mass terms of the matter fields are suppressed with that same
  factor $m_\mathrm{soft}\sim m_{3/2}/\langle X\rangle\sim  m_{3/2}/4\pi^2$. If
  we thus assume that the soft terms are in the region of the weak scale,
  $m_{3/2}$ will be in the multi TeV region and thus heavy as well. 
\end{itemize}
The general mass pattern of the scheme is thus determined by this little
hierarchy $\langle X\rangle = \log(\MPl/m_{3/2})\sim 4\pi^2$ with
\begin{eqnarray}
 m_T
 ~\sim~
 \langle X\rangle\, m_{3/2}
 ~\sim~
 \langle X\rangle^2\, m_\mathrm{soft}\;.
\end{eqnarray}

\subsection{Phenomenological aspects}

The above-mentioned mix of gravity and anomaly mediation, i.e.\ the `mirage
mediation' scheme allows, at least in substantial regions of the parameter
space, to retain the attractive features of these mediation mechanisms while
discarding the problematic aspects. The most important issues are the following:
\begin{itemize}
 \item Anomaly mediation has the notorious problem of negative mass squares
  for some matter fields, in particular for the sleptons. In \textbf{mirage mediation},
  the `gravity mediated' contribution can render the slepton mass
  squares positive thus leading to a consistent framework.
\item We have a partial solution of the \textbf{flavour problem}.
  First of all, anomaly mediation is flavour-blind and thus does not
  cause the usual flavour problems. If, in addition, all the fields
  live on the D7 branes we have a common scalar mass from the modulus
  mediation. This additional feature is not a result of the scheme
  itself, but a consequence of the assumption concerning the origin of
  matter fields. Nevertheless, it is worthwhile to stress that in
  mirage mediation the flavour problem get ameliorated, and that the
  scheme is flexible enough to allow for the implementation of a
  mechanism that solves the flavour problem.
 \item There is also a partial solution to the \textbf{SUSY CP-problem}
  coming from the special property of the superpotential \cite{Choi:1993yd}.
  That is, the phases of the $A$-terms and gaugino masses are aligned. 
  However, the extreme smallness of the various electric dipole moments might
  require further alignment of phases (see, e.g., \cite{Falkowski:2005ck}).
\item The scheme leads to a distinct pattern for the spectrum of the
  low-energy effective theory. For example, it has been observed that
  the spectrum exhibits a {\bf mirage unification} scale
  \cite{Choi:2005uz,Endo:2005uy}---i.e., the gaugino and scalar masses
  meet at an intermediate scale (an energy scale well below the GUT
  scale). However, this mirage unification scale does not correspond
  to a physical scale. It has also been argued that the partial
  cancellation of the RG evolution of the soft masses may ameliorate
  the SUSY fine-tuning problem \cite{Choi:2005hd,Kitano:2005wc}.
\begin{figure}[h] 
\centerline{\includegraphics[scale=0.9]{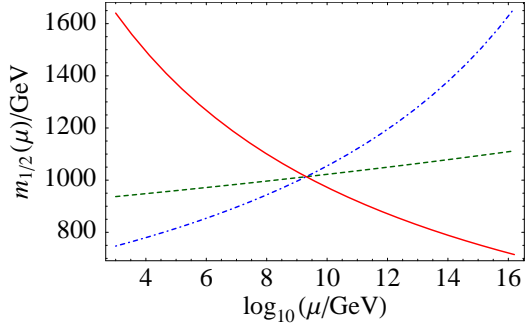}}
\caption{Mirage unification for $m_{3/2} = 40\, \mathrm{TeV}$, $M_0 :=F_T/T=
m_{3/2}/(4\pi)^2$ and $\alpha:=m_{3/2}/[M_0\times\log(\MPl/m_{3/2})]=1$. The red
(solid)/green (dashed)/blue (dash-dotted) curve shows the evolution of the
gluino/$W$ino/$B$ino mass.} 
\end{figure}
\item In contrast to most of the other schemes of SUSY breakdown, in
  mirage mediation the lightest superpartner (LSP) is dominated by
  the {\bf Higgsino} component in large regions of the parameter
  space~\cite{Endo:2005uy,Falkowski:2005ck,Choi:2005hd}.
\end{itemize}

\subsection{Cosmological aspects}

(Locally) supersymmetric theories are often in conflict with cosmology  because
they predict long-lived particles whose decays spoil the successful predictions
from nucleosynthesis. The most prominent examples for these long-lived particles
are the gravitino and the moduli. In the mirage mediation scheme, the latter are so
heavy that they decay early enough not to affect nucleosynthesis. This means
that the mirage mediation scheme does not suffer from the traditional gravitino
and moduli problems.

Let us mention that there are further challenges for moduli cosmology,
which persist even if moduli are rather heavy. These remaining
problems include: moduli may not find the minimum of their effective
potential at all; some of them might run to the phenomenologically
unacceptable run-away minimum \cite{Dine:1985he} due to a large
initial velocity \cite{Brustein:1992nk} or get destabilized by thermal
effects \cite{Buchmuller:2004xr,Buchmuller:2004tz}.  Nevertheless,
there exist a few promising proposals to solve at least some of these
problems (see, e.g.,
\cite{Horne:1994mi,Dine:2000ds,Kofman:2004yc,Brustein:2004jp}), but
these solutions may require some further ingredients.

\section{Summary}

We presented a scheme that combines the advantages of the new `flux
compactification' scenarios with the traditional lore of moduli
stabilization. Different from the usual models of `flux
compactification', a crucial feature of this scenario is that the
gaugino condensate does not adjust its size to a quantized constant.
Rather it sets the scale of SUSY breakdown, and thus yields the
explanation of the observed hierarchy $m_\mathrm{weak}\ll \MPl$
without the need of fine-tuning.  We argued that the `continuous
Chern--Simons term', which is comprised of `continuous Wilson lines',
should adjust its size such as to cancel the vacuum energy. In
particular, in this scheme there is no need for the (ad hoc) uplifting
procedure.

We further discussed the consequences of the scheme for phenomenology and
cosmology. Most importantly, there is the little hierarchy between moduli,
gravitino and soft masses, $m_\mathrm{modulus}\sim\langle X\rangle m_{3/2}\sim
\langle X\rangle^2 m_\mathrm{soft}$ with $\langle
X\rangle\sim\log(\MPl/m_{3/2})\sim4\pi^2$. The pattern of supersymmetry
breakdown combines the features of gravity/moduli and anomaly mediation. As we
have discussed, this leads to an attractive scenario where several problems
of supersymmetric extensions of the standard model are ameliorated or
even solved.





\bibliographystyle{aipproc}   

\bibliography{Moduli}

\IfFileExists{\jobname.bbl}{}
 {\typeout{}
  \typeout{******************************************}
  \typeout{** Please run "bibtex \jobname" to optain}
  \typeout{** the bibliography and then re-run LaTeX}
  \typeout{** twice to fix the references!}
  \typeout{******************************************}
  \typeout{}
 }

\end{document}